**Visualizing plasmon coupling in closely-spaced chains of Ag nanoparticles by electron energy loss spectroscopy**

By *Fengqi Song [1,3], Tingyu Wang [2], Xuefeng Wang [2], Changhui Xu [4], Longbing He [4],*

*Jianguo Wan [1], Chris Van Haesendonck [3], Simon P. Ringer [2*], Min Han [4*], Zongwen*

*Liu [2], Guanghou Wang [1]*


[*] Prof. S. P. Ringer and Prof. M. Han, Corresponding-Authors,
s.ringer@usyd.edu.au, sjhanmin@nju.edu.cn
The first two authors (F. Song, T. Wang) contribute equally.
[1] Dr. F. Song, Prof. J. Wan, Prof. G. Wang
National Laboratory of Solid State Microstructures and Department of Physics,
Nanjing University, Nanjing, 210093, P. R. China
[2] Dr. T. Wang, Dr. X. Wang, Dr. Z. Liu, Prof. S. P. Ringer
Australian Key Centre for Microscopy and Microanalysis, the University of Sydney,
Sydney, NSW2006, Australia
[3] Dr. F. Song, Prof. C. Van Haesendonck
Laboratorium voor Vaste - Stoffysica en Magnetisme, Kathelieke University of
Leuven, Leuven, B-3001 Belgium
[4] Prof. M. Han, Mr. C. Xu, Mr. L. He
National Laboratory of Solid State Microstructures and Department of Material
Science and Engineering, Nanjing University, Nanjing, 210093, P. R. China


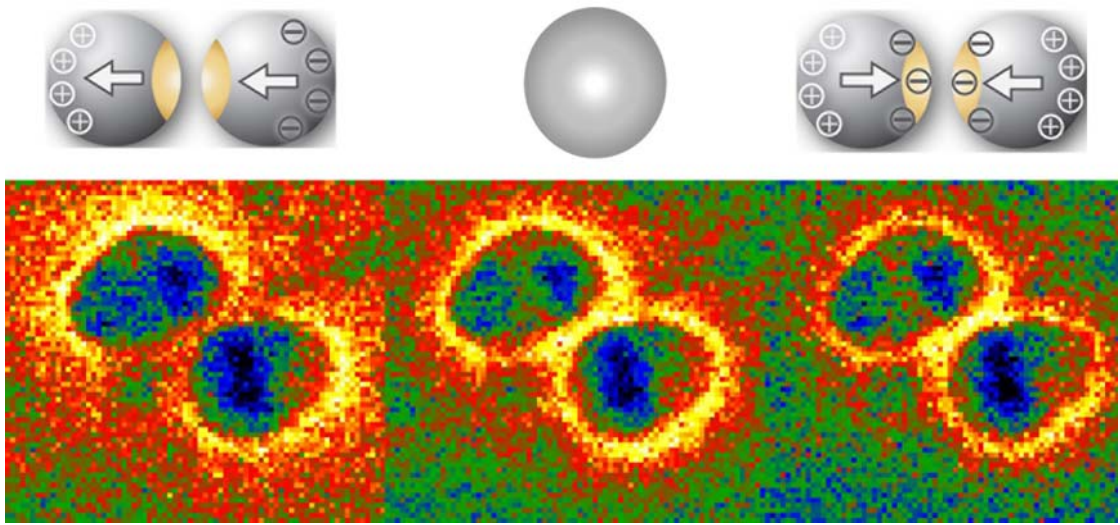





**ABSTRACT** Anisotropic plasmon coupling in closely-spaced chains of Ag nanoparticles was visualized using the electron energy loss spectroscopy in a scanning transmission electron microscope. For dimers as the simplest chain, mapping the plasmon excitations with nanometers' spatial resolution and 0.27 eV energy resolution intuitively identified two coupling plasmons. The in-phase mode redshifted from the ultraviolet region as the inter-particle spacing was reduced, reaching the visible range at 2.7 eV. Calculations based on the discrete dipole approximation confirmed its optical activeness, where the longitudinal direction was constructed as the path for light transportation. Two coupling paths were then observed in an inflexed 4-particle chain.







## 1. Introduction

Plasmon coupling in chains of noble metal nanoparticles has attracted much recent attention because the prospect of plasmon manipulation and it has potential applications in surface enhanced Raman scattering, nano-optics, nanoscale energy transportation and plasmonics/ photonic/nano-bio devices [1-12]. When the inter-particle spacing decreases below the wavelength of incident light, far-field interference ceases and gives rise to the near-field coupling. The near-field component sharply increases when the spacing reduces further to less than the diameter of an individual nanoparticle. Local field enhancement with the order of $10^3$ occurs between the nanoparticles [13-14] and even offers the opportunity to single molecule optics [11, 15]. Through the enhanced inter-particle coupling, light with selected wavelength can efficiently transport along the chain in some certain directions [2]. In this meaning, tracking the inter-particle plasmon coupling essentially reveals the paths of light transportation in the nanoparticle chains.

However, visualization of anisotropic coupling of inter-particle plasmons within chains of nanoparticles are experimentally challenging, especially for nanoparticles with the diameter and spacing <20 nm [9, 16-19]. Optical absorption spectra that are commonly used in the plasmon detections fail here. It is because the task involves anisotropic plasmon transportation in tens-of-nanometers, the detection of which requires, simultaneously, directional sensitivity, nanometer spatial resolution and high energy resolution and also relies on preparing nanoparticles with uniform spacing [20-21]. Firstly, optical probes observe the ensembles containing huge number of such





nanoparticles and chains within a light wavelength [16, 21]. Therefore, investigation of the plasmon coupling by optical absorption requires large-area lattice of identical chains with precise and uniform particle size and spacing, or locating a given chain with nanometers' precision [20, 21]. Sensitivity over the nanometers' spacing claims advanced preparation techniques [3, 5-7, 9, 13, 18, 20]. Secondly, many lattice points are included within a single light wavelength in as-prepared samples, which reasonably generates the coupling between the lattice points and some unphysical features [13, 18]. In such background, even the simplest case of plasmon coupling in dimers of nanoparticles is still in progress [1, 2, 10, 13, 16]. Finally and importantly, directional incidence of polarized light has been the optical method with the best directional sensitivity [7], but it still have difficulties in revealing the anisotropic transportation even if the lattice of identical points are successfully prepared, particularly for multi-particle chains with inflexions.

Here we combine cluster beam deposition (CBD) and electron energy loss (EEL) spectroscopy (EELS) in a cold field-emission scanning transmission electron microscope (STEM) in the task. Mapping surface plasmon on a single Ag nanoprism by STEM-EELS has recently proved intuitive observation of plasmon excitation [22], merging the plasmon studies by EELS to modern plasmonics [19, 23-24]. Similarly to the optical probe, the electron beam generates electric fields with multiple frequencies around the transmitting position [25]. Such electric fields may couple with the fields of intrinsic plasmon modes of the nanoparticles, which results in a response spectrum of plasmon excitation. People are connecting the plasmon





response by EEL excitation to optical absorption spectra [19, 23]. A great priority lies

that plasmon excitation by the electron beam is greatly localized and this imparts the

electron beam with the directional sensitivity on nanoscale. CBD is featured for

producing films with dispersed nanoparticles [26], where one can easily find

closely-spaced dimers with various spacing, even chains of multiple nanoparticles [20].

Such combination enables us to probe the plasmon coupling of individual chains with

nanometer's resolution and directional sensitivity. In the work, we first concentrate on

dimers of Ag nanoparticles. Anisotropic plasmon excitation is visualized and two

coupling plasmons are thus identified. The spacing-dependent behaviors of the

coupling plasmons are studied. Light interaction calculations based on the discrete

dipole approximation (DDA) confirm optical activeness of the in-phase plasmon and

also the possibility of light transportation through the path. Two coupling paths are

observed in an inflexed 4-particle chain.

## 2. Plasmon excitation in individual nanoparticles

Measurement of individual Ag nanoparticles validated the experimental routines

at first. A VG HB601 STEM was working with the energy resolution of 0.27 eV.

**Figure 1(a)** shows an EELS intensity map with the energy window of 3.3-3.6 eV and

the inset in Figure 1(a) is its STEM image. It is an individual Ag particle. The map

was obtained by integrating the intensities of the selected energy windows from EEL

spectra at the imaging points. Brightness of a point indicates the chance of plasmon

excitation when the electron beam is sampling on the point. We can find the intensity





maximums are distributed along the circular edge of the particle, where the electron beam excites the plasmon effectively. We can see the line profile of the plasmon intensities in Figure 1(a). The plasmon peak ceases rapidly within the distance of 2 nm. Such high localized excitation excludes the contribution from the none-of-interest nanoparticles, which confirms the unique priority of the electron beam over optical probes. It promises high resolution of the plasmon mapping and coupling by the electric field vectors also guarantees the directional sensitivity.

In Figure 1(b), we can find the spectra from the middle and the edge of the particle respectively. The plasmon maximum in the edge spectrum is located at 3.4 eV, which is assigned to the spherical mode of Ag nanoparticles. It is supported by the evidence from both circular maximum in plasmon mapping and fair consistence with the previous knowledge [22-23, 27-28], i.e. the peak at 365 nm in optical absorption spectra. The particles from the free CBD are often treated as spheres [26, 29]. When the electron beam is in the middle of the nanoparticle, another feature at 3.8 eV emerges due to the interband transition of the bulk Ag [23, 30]. Obvious identification of the two modes demonstrates present energy resolution of better than 0.3 eV. Figure 1(c) illustrates the positions of the spherical mode against the particle diameters. A steady redshift is seen and it reaches less than 3.3 eV for the 100 nm-diameter particle. But the position of the spherical plasmon remains rather steady from 10 to 30 nm, which rules out the influence of damping effects in the region [31].





## 3. Visualizing plasmon coupling in dimers of nanoparticles

The coupled plasmon modes were identified by mapping the plasmon of Ag dimers, in which anisotropic plasmon excitation was observed. For a given dimer with two identical spheres, there is a symmetrical axis and a center in the double-sphere frame [32]. For a highly localized probe, such basic anisotropicity expects distinct spectra from three regions as shown by the spots in **Figure 2**(b), i.e. the edge parallel to the center axis (blue-triangular), the edge perpendicular to the axis (black-box) and the center (red-circle).

The EEL spectra extracted from the spots turned out as expected in Figure 2(a). They were obtained from a 2 nm-gap-spacing dimer of two Ag nanoparticles, whose diameters were around 18 and 19 nm respectively. The spectrum taken from the blue-triangular spot resembles the spectrum of single particles as seen in Figure 2(a) and Figure 1(b). The spectrum from the black-box spot presents a new feature at 2.7-2.9 eV, whereas the spectrum from center (red-circle spot) presents a slight blue shift to 3.6 eV. Full mapping of the EEL spectra intensities from the selected energy windows (Figure. 2(c): 2.6-2.9 eV, 2(d): 3.2-3.5 eV, 2(e): 3.5-3.8 eV) provides unambiguous evidence. The maximum intensities of EEL spectra in (c), (d) and (e) fall around the black, blue and red spots respectively.

Plasmon excitation in EEL spectra has long been understood by coupling of electric fields from the incident electron and the nanomaterials [16, 18, 27]. Localized electric field from the incident electron beam can be simply decomposed in three directions of the double-sphere frame, i.e. the direction of the center axis, the incident





direction of electron beam and the third independent direction. The vector components in the latter two directions make no significant influence on the plasmon resonance of the neighboring nanoparticle since they are perpendicular to the axis. The vector component along the center axis acquired from the blue-triangular point also hardly couples to the neighboring particle, generating similar plasmon response to the individual ones. Remarkably, the component's excitation at the red-circle spot symmetrically pushes the dipoles of two particles and leads to an anti-phase dimer mode, while the excitation at the black-box spot favors an in-phase dimer mode with two in-phase dipoles as shown by the insets of Figure 3(b). This assigns the in-phase mode at 2.7-2.9 eV and the anti-phase mode at 3.6 eV. The positions agree to the theory [14].

The coupling modes are sensitively manipulated by the inter-particle spacing. The experimental data in Figure 3(b) shows that no observable shift or new features emerge when the spacing is comparable to the particle's diameter (ratio=1). The shifts become visible with the ratio of less than 0.2. The in-phase mode continuously redshift to as low as 2.7 eV at the spacing of 1.5 nm, while the anti-phase mode exhibits a slight blue shift. Note that the coupling distance between the plasmon of Ag particles is even comparable to their diameters, while the coupling distance between the electron probe and the nanoparticle plasmon is much smaller. It reveals different physical origins of the two couplings [33].





## 4. Correlating the plasmon excitation by electron beam to optical incidence

Plasmon response excited by electron beam can be optically active if it is a dipolar resonance [23]. Electron beam generates an electric field function with contributions of all frequencies [24-25, 27]. All the components with different frequencies perturb the particle plasmon and yield a spectrum of plasmon response. The maximum excitation is expected close to the frequency of free plasmon resonance [19]. Therefore, we found the maximum excitation at 3.4 eV both in Figure. 1(b) and the calculated optical extinction spectrum of an individual Ag nanoparticle.

The observed in-phase plasmon is a dipolar mode and eligible for coupling with the incident light. It appeared in optical extinction spectra by DDA calculations of light interaction in **Figure 3** and its spacing-dependent behavior was also reproduced. Figure 3(a) is the calculated extinction spectra based on the double-sphere model. A similar spectrum to that of single particles presents a single peak at 3.43 eV at the ratio of 1 (Figure 3(a) and Figure 1(a)), where plasmon coupling generates no fresh feature. The peak splits and the low-frequency mode appears when the ratio is decreased to 0.4. It continues to redshift along the dashed line with decreasing gap spacing and reaches less than 3 eV at the ratio of 0.05. One may see the other two obvious features at 3.2 and 3.6 eV in the spectrum (ratio=0.05), which are from the contribution from high-order modes and also appear in the black experimental curve of Figure 2(a) [14]. The black curve in Figure 3(b) shows the redshift of the in-phase mode with decreasing gap spacing. A sharp redshift presents in the nearly-touching





region. All the evidence supports that the experimentally-detected in-phase coupling mode is eligible for coupling with light, therefore light can be transported along the longitudinal direction of the dimers by the redshifted plasmon. We essentially visualize the transportation path of incident light.

**5. Visualizing coupling paths in an inflexed chain of nanoparticles**

Knowledge in dimers' plasmon contains basic ideas of plasmon coupling in nanoparticle chains. As stated above, the electron beam is a nano-localized exciting probe of plasmon responses, which therefore traces the transit plasmon modes by anisotropic plasmon excitations. Firstly, it might generate multipolar resonance, which might be invisible in optical absorption [14, 22]. For example, a mode in the green-box curve is invisible in the calculated extinction spectrum (**Figure. 4**(c)). It was extracted from the center of three nanoparticles, containing the contribution of a tripolar mode as shown by the green-box spot in Figure 4(a) [22, 34], while such resonances are hardly coupled to incident light wave. Secondly, we can see some smaller particles with the diameters of 3-6 nm. The gap spacing between the small particles and the 4 dominant ones are comparable or even larger than the diameters of small particles. They were found to only transport the plasmon resonance of large particles without generating new modes. This concludes little influence to the resonant frequencies of large dominant particles from them [7, 24, 34].

Finally, anisotropic excitations are also shown in Figure 4(c) although ideal spatial mappings (like Figure 1(a) and Figure 2(c), (d), (e)) can't be obtained due to





mixing more complex modes than dimers and blurred spatial distributions of the intensities by the small particles. There could be even more than 1 mode in the range of 0.27 eV, i.e. within the resolution of the method sometimes. Based on the understanding of dimers, the blue-triangular spot prefers to excite an in-phase coupling mode of dimers, which is at 2.7 eV as seen in the blue-triangular curve of Figure 4(c). A multi-order mode at 3.0 eV is also found. The red-circle spot will favor a mode due to in-phase coupling trimers, which is expected in lower frequency by calculation in the top of Figure 4(c) [3, 30]. It is at 2.3 eV in the red-circle experimental spectrum Figure 4(c). Main distinction between the spectra from the short-chain point and long-chain point is the rise of the low-frequency mode, which also appears in the calculated extinction spectrum (black smooth in Figure 4(c)). The revealing of the transit plasmons visualizes the two coupling paths for light transportation in the inflexed particle chain. Positions of the transit plasmon mode redshift when the particle number is increasing. It is 3.4 eV for single particle, 2.7 eV for a dimer and decreases to around 2.3 eV for a chain of three nanoparticles.

## 6. Conclusion

Plasmon coupling through chains of Ag nanoparticles with the diameters and inter-particle spacing of less than 20 nm was investigated by STEM-EELS technique. In the study of dimers, the in-phase and anti-phase coupling modes were identified by anisotropic plasmon mapping. The in-phase mode redshifted with decreasing dimer gaps. DDA calculations based on the double-sphere model demonstrated the optical





activeness of the in-phase coupling transit mode. This constructs the longitudinal direction as a coupling path, through which light can be transported. Two coupling paths are visualized in an inflexed 4-particle chain.

### 7. Experimental

The particle beam was generated by a gas-aggregation cluster source described elsewhere [20]. The particle sizes were varied from a few to even 100 nanometers. The beam was deposited onto the TEM copper meshes with an ultrathin film of amorphous carbon. No obvious features were found from the carbon film at least in the interested region from 2 to 4 eV. The EEL spectra were collected by a VG HB601 STEM with a cold field-emission gun. It was working at 100 keV and the full width at half maximum (FWHM) of the zero-loss peak (ZLP) was tuned to 0.27 eV. The probing energy region was from -5 eV to 27.5 eV with the energy dispersion of 0.025 eV. The probing step was 0.5 nm. Both the intervals allowed accumulation of binned points to achieve the spectra with the satisfied signal-to-noise ratio. The probing period at a single point was set to 0.1 s to limit the influence of space drift. In the data processing, the spectra were smoothed by energy window of 0.2 eV to remove unphysical features. Richardson-Lucy algorithm of less than 3 iterations was further implemented [35-36]. Most of the processing was carried out by Gatan Digital Micrograph. We note the observed nanoparticles are a little distorted from the ideally symmetrical shapes we assume in the present interpretations. Here it doesn't make obvious errors since previous plasmon calculations of an ellipsoid provide the





deviation from the plasmon positions of spheres, which monotone depends on its distortion.

DDA Calculations of the extinction spectra were implemented by DDSCAT7.0 [37]. The whole space was simulated by the grids of $100 \times 100 \times 100$, where Ag nanoparticles were arranged according to the experiments. Typically, 1000-5000 dipoles were generated within a sphere. The dielectric function of the Ag nanoparticles was adapted from the experimental values [28, 30, 38-39], therefore, only 30 points were available in the interested region between 2 and 4 eV. Considering the current experimental energy resolution of 0.27 eV, the calculated spectra are acceptable for the comparison. For a given dimer, the light incidence was perpendicular to the axis. We used the dimer with two equal spheres because the two particles were similar in diameters experimentally. Calculations for the 4-particle chain are described below. Coupling through carbon film was not considered because it is a tripolar mode [25].

*Acknowledgements*
This work was financially supported by the Discovery Project of Australian Research Council, the National Natural Science Foundation of China (Grant Nos. 90606002, 10674056, 10904100, 10775070), the National Key Projects for Basic Research of China (Grant No. 2009CB930501，2010CB923401), and the Program for New Century Excellent Talents in University of China (Grant No. NCET-07-0422). The Australian work is supported by the Australian Microcopy & Microanalysis Research Facility (AMMRF) at the Australian Key Centre for Microscopy and Microanalysis at the University of Sydney. We thank Mr. Shaun Block, Dr. Michel Bosman and Dr. David Mitchley for valuable discussions.

**Figure Legends**





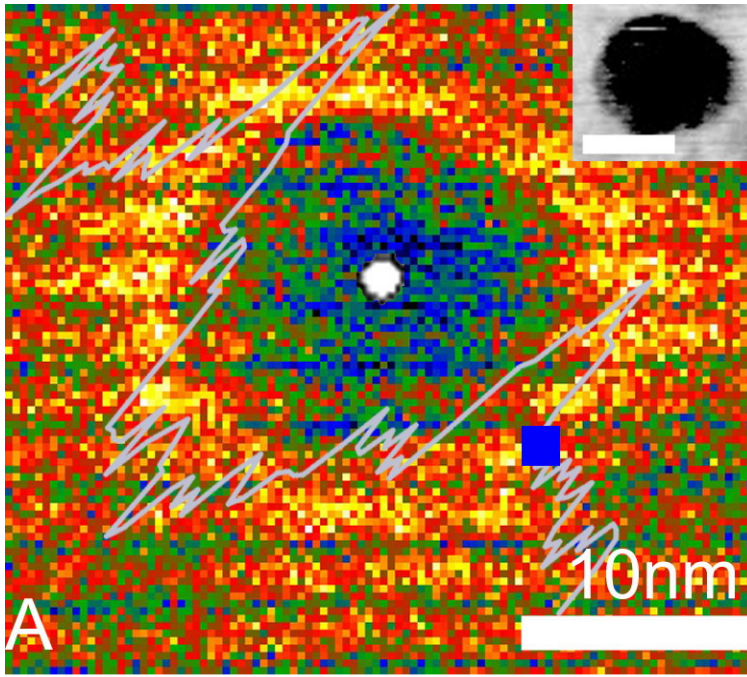

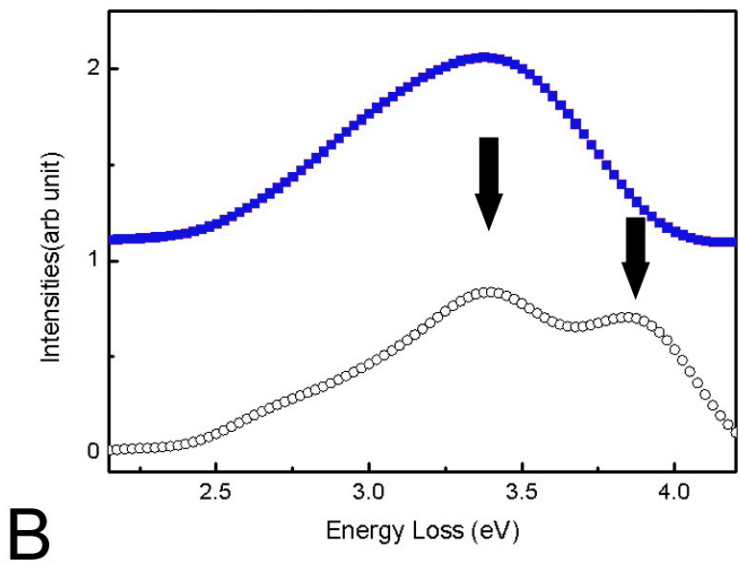

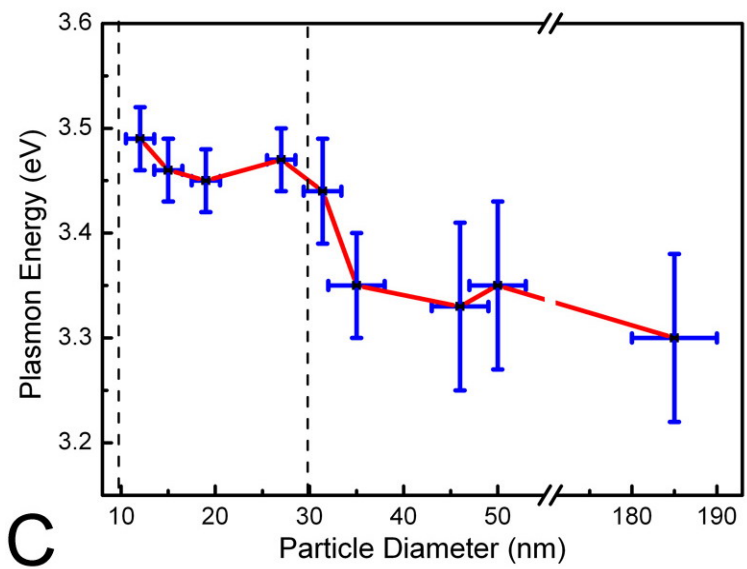



**Figure 1**. Probing individual Ag nanoparticles. (a) EEL intensity mapping of an Ag particle (energy window: 3.3-3.6 eV). The grey curve is its line profile and the top right is its STEM image. Two spots mark the extracting positions of EEL spectra in (b). (b) EEL spectra from the spots in (a). The top curve is from the white-circle spot and the bottom curve is from the blue-box spot. The black arrows mark the positions of spherical mode at 3.4 eV and bulk mode at 3.8 eV. Features with the FWHM less than 0.2 eV are unphysical. (c) Positions of the spherical mode with cluster diameters. More particles are from 10 to 30 nm because of size distribution of the beam, leading to smaller error bars. The position keeps stable from 10 to 30 nm.





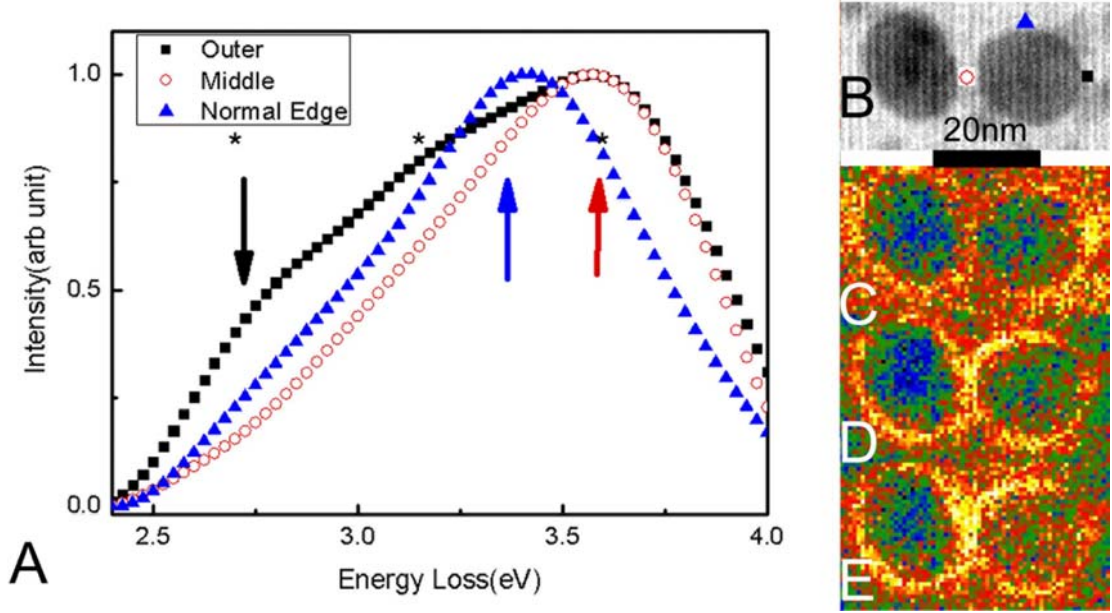

**Figure 2**. Probing a dimer of silver particles. (a) EEL spectra from the spots marked in the STEM image of (b), where the red-circle, black-box and blue-triangular curves refer to the middle, the outer edge and the perpendicular edge of the cluster dimer respectively. The black asterisks marked three features of the spectrum from the black spot, which are consistent to the calculated spectrum in Fig 3a. (c), (d) and (e) are the EEL intensity mapping with the energy windows of 2.6-2.9, 3.2-3.5 and 3.5-3.8 eV, respectively. The brightest dots mean the intensity maximum in the energy windows. The mapping shows the plasmon mode at 2.8 eV is excited with the largest chance along the outer edge (black-box spot), the maximum excitation of 3.6 eV is positioned in the middle of the dimer (red-circle spot). The same situation is shown more clearly in the toc figure. Current energy resolution generates poor separation between the modes at 3.4 and 3.6 eV, leading to the moderate contrasts in (e).





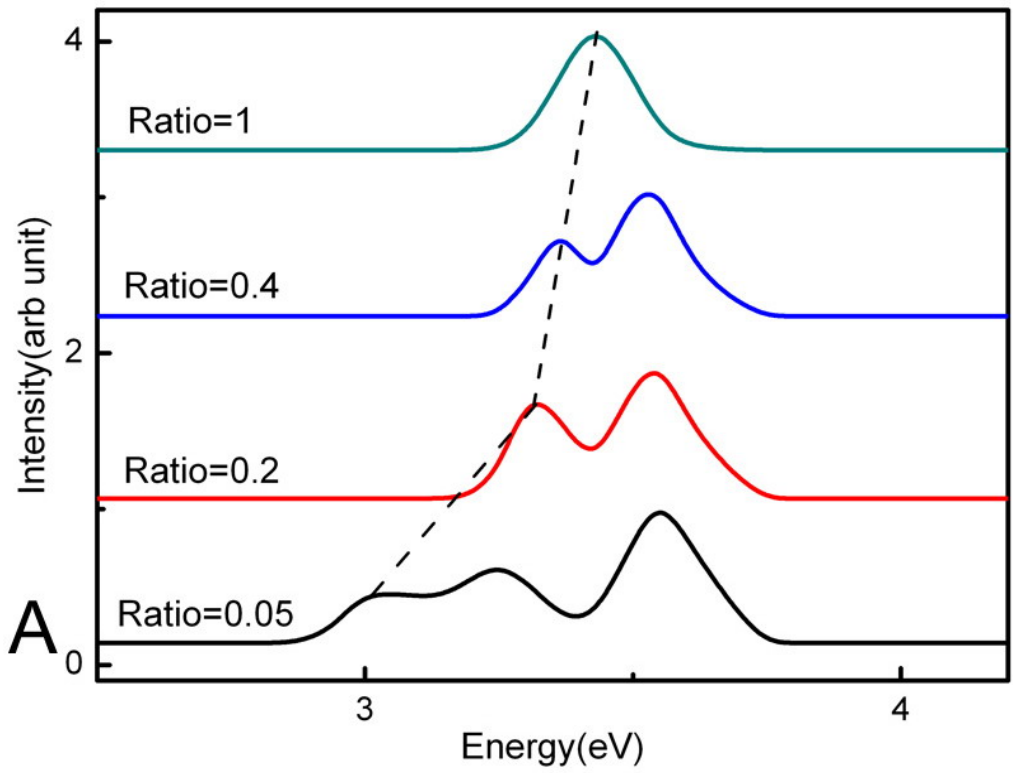

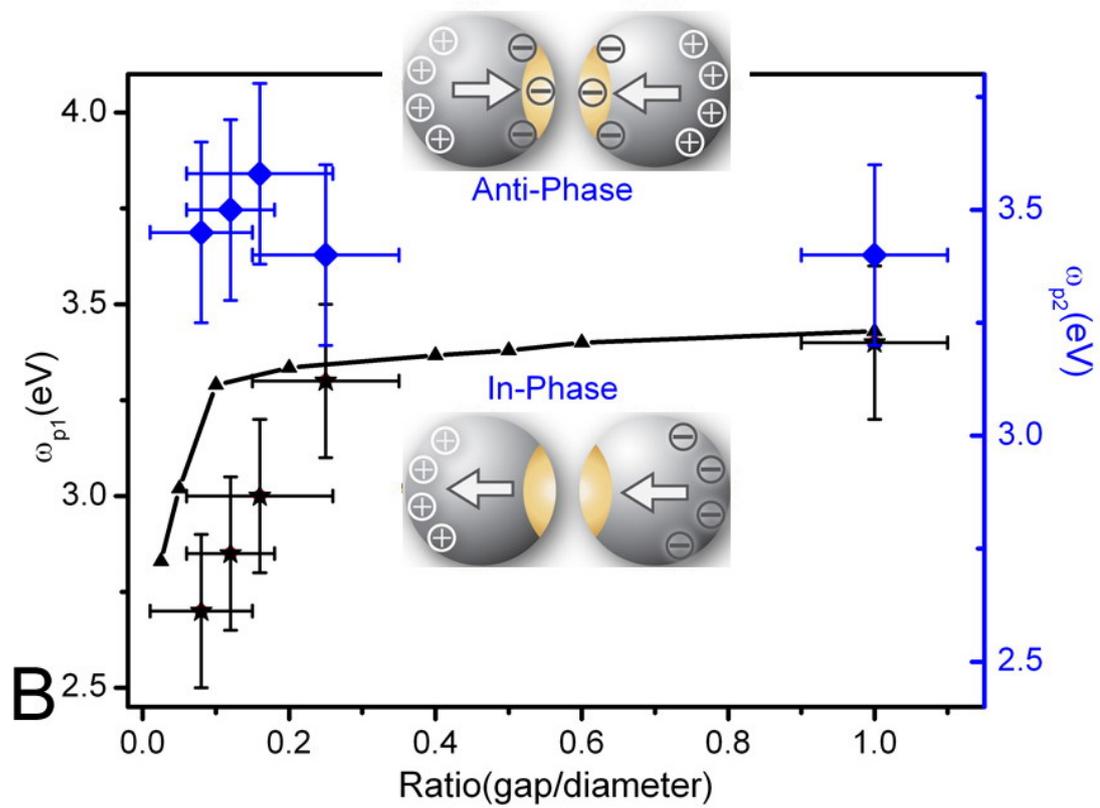





**Figure 3**. Modeling by DDA calculations and comparisons with the experiments. (a) are the extinction spectra from the DDA calculation based on the double-sphere model as shown in the inset of (b). The dashed line in (a) marks the shift of the in-phase modes when the particles are approaching. The incident light is perpendicular to the center axis. In (b), the anti-phase mode and the in-phase mode are shown in blue and black respectively. The black curve plots the calculated positions of the in-phase mode. The insets illustrate the charge resonance of the two modes. Dots with error bars are from the experimental data. We can see the consistence of the trends between the calculation and the experiment in (b). We also made calculation based on the cap-ending cylinder model, which is often used considering the spheroids and nanorods[6, 24, 32]. However, the double-sphere model better reproduced the experimental results. It further validates the sphere approximation in current Ag particles.





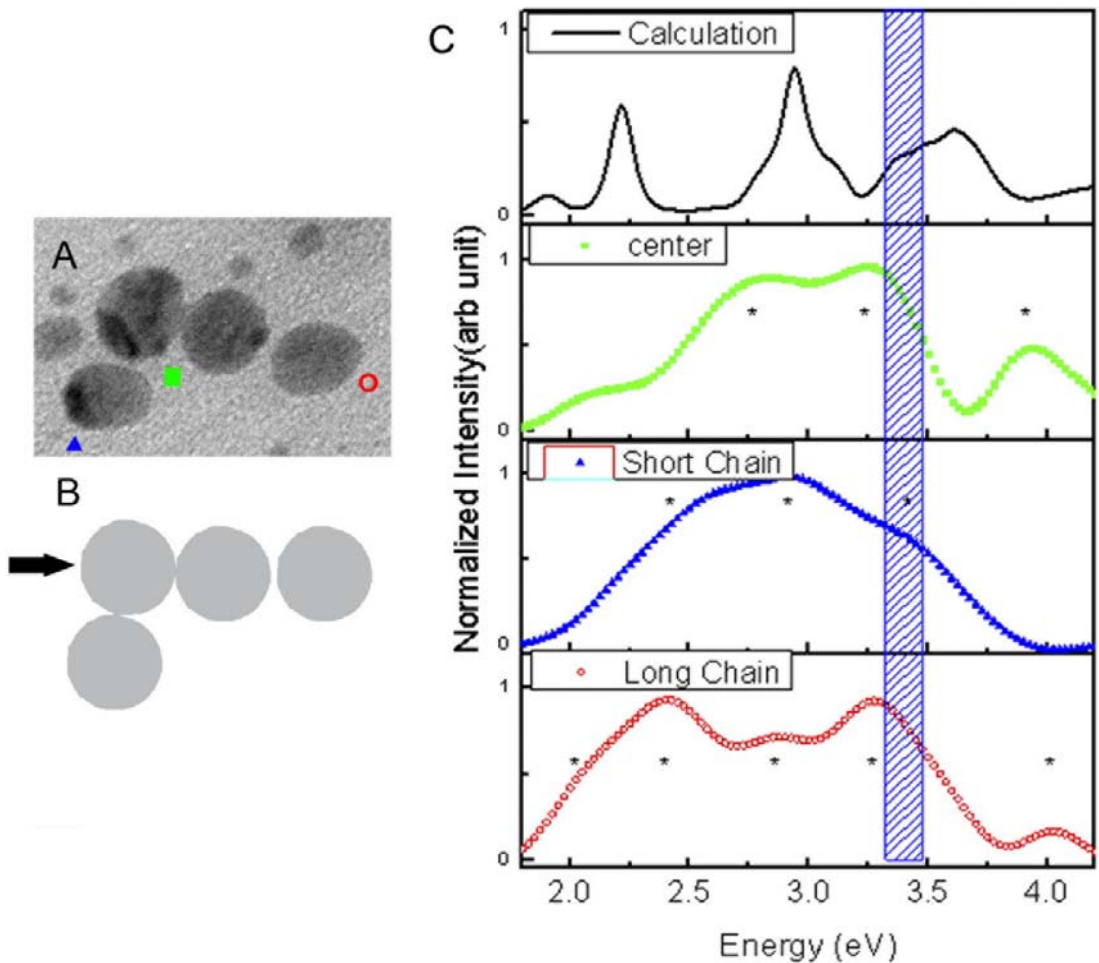

**Figure 4**. Measuring a chain of nanoparticles. (a) is its STEM image, where the red-circle, blue-triangular and green-box spots mark the positions where the spectra are extracted, (i.e. along the long chain, short chain and in the center respectively). The spots correspond to the spectra marked by the same spots respectively in (c). (b) is the model used in the calculation, which contains four identical spheres with diameters of 15 nm and the measured gap spacing. The black arrow shows the incident direction of light. (c) shows the spectra acquired from the spots with the corresponding colors in (a) and the calculated spectra according the configuration of (b). Due to the asymmetric ZLPs, the peaks are not deconvolved in Gaussian or other analytic forms, which causes inadequate contrast between different modes in plasmon





mapping.    Here the positions with visible features are marked by asterisks. The blue

bar covers the contribution of singular spherical mode at 3.4 eV.

**The table of contents entry**.

Here we show the mapping of plasmon excitations in a dimer of Ag nanoparticles, where the selected energy windows are 2.6-2.9 eV, 3.3-3.5 eV and 3.5-3.8 eV respectively. In the left inset, we can see the plasmon excitation is maximized along the outer edge of the dimer. The plasmon is demonstrated to be an in-phase coupling mode and optically active.


Fengqi Song , Tingyu Wang, Xuefeng Wang, Changhui Xu, Longbing He, Jianguo Wan, Chris Van Haesendonck, Simon P. Ringer, Min Han, Zongwen Liu, Guanghou Wang


**Visualizing plasmon coupling in closely-spaced chains of Ag nanoparticles by electron energy loss spectroscopy**